\documentclass[sigconf]{acmart}
\usepackage{comment}
\newcommand*{\SuperScriptSameStyle}[1]{%
  \ensuremath{%
    \mathchoice
      {{}^{\displaystyle #1}}%
      {{}^{\textstyle #1}}%
      {{}^{\scriptstyle #1}}%
      {{}^{\scriptscriptstyle #1}}%
  }%
}
\newcommand*{\oneS}{\SuperScriptSameStyle{*}}
\newcommand*{\twoS}{\SuperScriptSameStyle{**}}
\newcommand*{\threeS}{\SuperScriptSameStyle{*{*}*}}
\usepackage{multirow}
\usepackage[justification=centering]{caption}
\usepackage{ragged2e}
\usepackage{natbib}
\usepackage{booktabs}
\usepackage{siunitx}
\usepackage{multirow}
\usepackage{colortbl}
\usepackage{hyperref}
\usepackage{graphicx}
\usepackage{framed}
\usepackage{pgf}
\usepackage{tikz}
\usepackage{tikz-dependency}
\usepackage{soul}
\usepackage[utf8]{inputenc}
\usetikzlibrary{arrows,automata}
\usetikzlibrary{positioning}
\usepackage{subcaption}
\usepackage{mwe}
\AtBeginDocument{%
  \providecommand\BibTeX{{%
    \normalfont B\kern-0.5em{\scshape i\kern-0.25em b}\kern-0.8em\TeX}}}

\copyrightyear{2022}
\acmYear{2022}
\setcopyright{rightsretained}
\acmConference[FAccT '22]{2022 ACM Conference on Fairness, Accountability, and Transparency}{June 21--24, 2022}{Seoul, Republic of Korea}
\acmBooktitle{2022 ACM Conference on Fairness, Accountability, and Transparency (FAccT '22), June 21--24, 2022, Seoul, Republic of Korea}\acmDOI{10.1145/3531146.3533222}
\acmISBN{978-1-4503-9352-2/22/06}
\begin{document}

%%
%% The "title" command has an optional parameter,
%% allowing the author to define a "short title" to be used in page headers.
\title[A Contextual Integrity Perspective on the Appropriateness of COVID-19 Vaccination Certificates]{Stop the Spread: A Contextual Integrity Perspective on the Appropriateness of COVID-19 Vaccination Certificates}
%%
%% The "author" command and its associated commands are used to define
%% the authors and their affiliations.
%% Of note is the shared affiliation of the first two authors, and the
%% "authornote" and "authornotemark" commands
%% used to denote shared contribution to the research.
\author{Shikun Zhang}
%\authornote{Both authors contributed equally to this research.}
%
%\orcid{1234-5678-9012}
\affiliation{%
  \institution{Carnegie Mellon University}
  %\streetaddress{P.O. Box 1212}
  \city{Pittsburgh}
  \state{PA}
  \country{USA}
  \postcode{15213}
}  \email{shikunz@cs.cmu.edu}

\author{Yan Shvartzshnaider}
\affiliation{%
  \institution{York University}
  %\streetaddress{1 Th{\o}rv{\"a}ld Circle}
  \city{Toronto}
  \country{Canada}}
\email{yansh@yorku.ca}

\author{Yuanyuan Feng}
\affiliation{%
  \institution{University of Vermont}
  \city{Burlington}
  \state{VT}
  \country{USA}
  \postcode{05405}
}\email{yuanyuan.feng@uvm.edu}

\author{Helen Nissenbaum}
\affiliation{%
 \institution{Cornell Tech}
 %\streetaddress{Rono-Hills}
 \city{New York}
 \state{NY}
 \country{USA}
 %\postcode{XXXXX}
 }
 \email{hn288@cornell.edu}

\author{Norman Sadeh}
\affiliation{%
  \institution{Carnegie Mellon University}
  %\streetaddress{P.O. Box 1212}
  \city{Pittsburgh}
  \state{PA}
  \country{USA}
  \postcode{15213}
}
\email{sadeh@cs.cmu.edu}

%%
%% By default, the full list of authors will be used in the page
%% headers. Often, this list is too long, and will overlap
%% other information printed in the page headers. This command allows
%% the author to define a more concise list
%% of authors' names for this purpose.
%\renewcommand{\shortauthors}{Zhang, Shvartzhnaider, Feng, et al.}

%%
%% The abstract is a short summary of the work to be presented in the
%% article.
\begin{abstract}
  We present an empirical study exploring how privacy influences the acceptance of vaccination certificate (VC) deployments across different realistic usage scenarios. The study employed the privacy framework of Contextual Integrity, which has been shown to be particularly effective in capturing people's privacy expectations across different contexts. We use a vignette methodology, where we selectively manipulate salient contextual parameters to learn whether and how they affect people's attitudes towards VCs. We surveyed 890 participants from a demographically-stratified sample of the US population to gauge the acceptance and overall attitudes towards possible VC deployments to enforce vaccination mandates and the different information flows VCs might entail. Analysis of results collected as part of this study is used to derive general normative observations about different possible VC practices and to provide guidance for the possible deployments of VCs in different contexts. 
\end{abstract}

%%
%% The code below is generated by the tool at http://dl.acm.org/ccs.cfm.
%% Please copy and paste the code instead of the example below.
%%

%%
%% Keywords. The author(s) should pick words that accurately describe
%% the work being presented. Separate the keywords with commas.

%%
%% This command processes the author and affiliation and title
%% information and builds the first part of the formatted document.
\maketitle

\section{Introduction}

The prolonged and devastating COVID-19 pandemic has affected every aspect of people's lives as well as the global economy. In an attempt to curb the spread of highly contagious variants, governments around the world have contemplated or adopted vaccination mandates (VMs) and vaccination certificates (or passports) (VCs) in schools, hospitals, public transportation, and other social contexts~\cite{IsraeliWorkplace,CollegeMandate,CanadaVaxpass,EUGreenCert,Excelsior,HawaiiVC,BritainBarVCPassport}. COVID VMs and VCs challenge established societal norms and conventions. While vaccination mandates and certificates are not new (e.g., vaccination mandates for children attending schools, ``yellow cards'' for travel to or from a country with a high risk of diseases such as yellow fever~\cite{yellowcard}),
the sudden and unprecedented requirement to show proof of vaccination to gain access to public venues or engage in a range of daily activities has triggered a fierce global debate on the appropriateness of COVID-19 VMs and VCs in light of established societal norms and conventions, perceived privacy harms, and civil liberty expectations~\cite{SupremeNews,APAustrian,GermanMandate,Kofler2021TenReasons,WEForumVMDebate}. 

Some proponents of VMs and VCs argue for overriding these values in the short term to accommodate urgent public health needs. Those opposing these measures warn against grave long-term repercussions that could result from bending traditional norms and civil liberty expectations. The arguments of both sides have merits. On the one hand, with over 6 million COVID-related deaths worldwide at the time of writing~\cite{WhoCOVIDDashboard}, enforcing VMs and requiring VCs is compelling, with data showing that vaccinations have a proven record of reducing the number of infections and hospitalizations~\cite{CDCVaccineEffective}. On the other hand, without proper policies and technology-backed measures, the collection and use of the information contained in VCs, such as an individual’s ID number, full name, date of birth, gender, nationality, and vaccination records, may not be restricted to its intended context or purposes. Risks of the digitization and re-purposing of this information have prompted warnings against large-scale adoption of VCs~\cite{IPThreatPrivacy, ExpertPrivacy2021} that could result in privacy violations, widening inequalities, and discrimination~\cite{Cofone2021,Baylis2021,Liz2021}.

In our view, the privacy implications of VC-related and other technologies depend on the context in which they operate. For example, it might be appropriate for vaccination status and date of birth to be shared with a doctor in a health context for a medical evaluation, while not so, when the same information is shared with an event organizer to gain access. Against the background of the global pandemic and increasing adoptions of VMs and VCs, it is important to ensure that the information flow rules embodied in the new technologies adhere to prevailing societal norms for each given context. %For instance, while some people are likely to find VC requirements acceptable to board a flight but perhaps excessive for booking a hotel room or trying to rent an apartment. 
The present study was conducted to develop a deeper understanding of what the respective societal norms are and whether the information flows that VC deployments give rise to are appropriate, with the help of the Contextual Integrity (CI) framework~\cite{nissenbaum2009privacy}.

%,to examine and re-evaluate whether the VC information flows breach established information governing norms and institutions. 
%and can be addressed in the design and deployments of VC-related technologies. 
%we need to u

%The present study was conducted to develop a deeper understanding of the appropriateness and legitimacy of information flows that VC deployments give rise to.
%In our work, we use the Contextual Integrity (CI)~\cite{nissenbaum2009privacy} framework to evaluate the privacy implication of VC technologies in terms of appropriateness and legitimacy of information flows they give rise to. 
We use an established CI-based vignette survey methodology~\cite{apthorpe2019evaluating,Shvartzshnaider2016} to analyze data from a US-based demographically-stratified sample ($N=890$) about how they perceive sharing VC information with various recipients in different contexts such as education, health, or public transportation, under different conditions and for various purposes. %This study was conducted in July 2021, providing a detailed snapshot of people's attitudes towards the adoption of VCs about 16 months into the pandemic. 
Our analysis reveals that perceived appropriateness is contextual and varies depending on CI's five parameters for information flows (i.e., sender, attribute, subject, recipient, transmission principle). There is also a significant difference in acceptance of first-hand information sharing compared to later re-sharing and re-purposing of originally collected information. Overall, we find that information re-sharing with entities other than public health agencies is widely viewed as unacceptable. These findings are relevant both to policy and technology design because they may reveal that even though VCs alter information flows, not all of these alterations constitute violations of privacy; benefits may be enjoyed without---in these instances---a need to curtail societal values.
\section{Background and Related Work}
In this section, we describe how this study builds upon related work around the topic of certifying immunity, recent user research on COVID-19 pandemic mitigation technologies, and the privacy framework of Contextual Integrity.

\subsection{Social and Privacy Considerations for Certifying Immunity}

As COVID-19 vaccines are increasingly available, several nations and industries have begun to develop digital solutions to certify people's immunity towards COVID-19~\cite{IsraelPass,EUGreenCert,Excelsior}. These efforts draw on the existing paper-based proof of vaccination counterparts~\cite{WHO1969} for diseases like smallpox, typhus, and cholera~\cite{VaccinePassports85Years}.
Although these solutions vary in what they certify (e.g., the presence of an antibody, a negative virus test, or a vaccination record), they introduce some digital verification mechanisms that may restrict people's access to different social activities or venues, such as traveling, attending large public events, or entering restaurants or bars. 
Deploying these solutions at scale has profound social implications: They could widen the existing inequities in access to healthcare resources and technologies; they could also lead to increased and systemic discrimination, especially for the already vulnerable~\cite{Baylis2021,baylis2020covid,Brown2021,Brown2021Reply,Phelan2020,Kofler2021TenReasons}. These risks have prompted the World Health Organization to discourage the use of VCs for international travel~\cite{WHOVaccineCertificates}.

More relevant to our work, mechanisms to certify one's vaccination status also potentially increase privacy risks. In addition to vaccination records, a typical VC includes an individual's national ID (e.g., passport number), full name, date of birth, gender, and nationality, most of which is personally identifiable information~\cite{yellowcard,IsraelPass,CanadaVaxpass}.
Although both paper and digital VCs ostensibly have similar features, digital VCs introduce new privacy challenges.
Digital information is inherently easier to collect and share on a much greater scale, possibly outside its intended context. For example, if people need to show their VCs to gain access to public venues and social events such as swimming pools, gyms, and concerts~\cite{IsraelPass}, data about their daily activities may be easily tracked and open to potential misuse. To gain a deeper and more comprehensive understanding of how deploying VCs and enforcing VMs could affect privacy and social life, our work explores the potential privacy violations of VC usage across a number of different contexts, which also enriches the discussion around the social implications of VMs and VCs.
\subsection{Surveying Public Opinions about Vaccination Certificates}
The use of technology in COVID-19 pandemic mitigation and containment, such as digital contact tracing apps, has spurred a number of studies into public opinion about these solutions, whose success relies upon broad user adoption~\cite{AustralianCOVIDTracking2021Garrett, williams2021public,o2021national, lohar2021irish, altmann2020acceptability,HowGoodCOVIDApps2020}.
Although these studies have revealed generally positive attitudes towards some of these solutions~\cite{lohar2021irish,williams2021public}, people often hesitate to install these apps out of concerns about cybersecurity and privacy~\cite{altmann2020acceptability, lohar2021irish}, as well as greater surveillance by governments and big technology companies after the pandemic~\cite{altmann2020acceptability, williams2021public}.

%ZhangB2020Covid (framework for VC
With increasing COVID-19 vaccination rates around the world, a number of non-academic surveys have attempted to gauge public attitudes towards and concerns around potential mechanisms, such as VCs, to verify people's vaccination status against COVID-19~\cite{QualtricsAmerican2020Haney, SwissImmunitySerology2020Nehme, BritonsImmunityCert2020Redfield, GermanyImmunitySnapshot2020, HillHarris}. 
In Germany, 45\% of the population was reported to oppose the introduction of VCs~\cite{GermanyImmunitySnapshot2020}. In Australia, 75\% of Australians were reported to support the use of VCs with only ~10\% opposing it~\cite{AustralianCOVIDTracking2021Garrett}. High levels of approval were also reported in Switzerland and in the UK, 60\% of Swiss~\cite{SwissImmunitySerology2020Nehme} and 69\% of Brits indicated they supported VCs~\cite{BritonsImmunityCert2020Redfield}.
A survey of the US population in April 2021 has found a mixed-level of support for using VCs: 53\% of Americans expressed support for government-issued vaccine passports, while 47\% reported being against the use of vaccine certificates~\cite{HillHarris}. 
Although the Brookings Institute's report~\cite{ZhangBB2021Brookings} outlining principles to build robust and ethical vaccination verification systems is an important contribution, nevertheless, comprehensive and rigorous academic research to supplement such contributions with insights on attitudes and conditions of acceptance behind potential support for the use of VCs is much needed. 

In our work, we aimed to take a rigorous and nuanced approach to understand the privacy concerns around VCs in their different contexts of use. Guided by the theory of Contextual Integrity, we surveyed a representative sample of the US population and studied their normative perceptions of sharing VC information with different entities, for varying purposes, and in diverse conditions.

\subsection{Studying Privacy through Contextual Integrity}\label{sec:CI_def} 

The theory of Contextual Integrity (CI)~\cite{nissenbaum2004privacy,nissenbaum2009privacy} provides a practical way to study privacy and assess the ethical implications of data handling practices. CI defines privacy in terms of the appropriate and legitimate flow of information. Appropriate flow, generally, is a function of conformance with established contextual norms, which are expressible in terms of five CI parameters: three actor parameters (\verb|sender|, \verb|recipient|, \verb|information subject|), an \verb|attribute| parameter, specifying the type of information, and the \verb|transmission principle| parameter, constraining the conditions under which information flows. Being able to specify the values for all 5 parameters is imperative to evaluating the privacy implication of any practice involving information flows. CI posits that a potential privacy violation occurs when one, or more of the information flow parameters, deviates from an established norm. For example, it might be considered appropriate for a store owner (recipient) to collect VC information (attribute) from a customer (sender) before letting them into the store (transmission principle). However, if the business owner were to collect this information for advertising purposes or get the VC information from a third party, the resulting flow---with a different transmission principle and sender---would deviate from the established expectation. According to CI, a deviation such as this may be experienced as a norm violation; in turn norm violations raise a red flag, signaling the possibility of a privacy violation. Although a complete analysis of the ethical implications of privacy norm violations requires a comparative philosophical assessment of norms versus novel flows, for the studies we report on, here, our focus is on people's judgments of appropriateness (or, people's privacy expectations.)

%Using Nissenbaum's contextual integrity (CI) privacy framework, we can evaluate the appropriateness of data practices of new technologies by considering important contextual factors~\cite{nissenbaum2004privacy,nissenbaum2009privacy}. Under the framework, data practices can be evaluated against established, contextual information norms using five CI parameters (i.e., the \verb|sender|, the \verb|recipient|, the \verb|attribute|, the \verb|subject|, and the \verb|transmission principle|.)

Previous efforts~\cite{Apthorper_2018,apthorpe2019evaluating,Shvartzshnaider2016,Martin_Nissenbaum_2015,zhang2021did,ZhangSOUPS2021} aimed at discovering privacy expectations have used the CI norm structure to inform vignette studies for investigating privacy implications of particular technologies in different contexts.
These studies have also shown that varying or omitting any of the five CI parameters has a significant effect on subjects' judgments of the appropriateness of particular information flows. 
Among studies using the CI framework, some have applied it to investigate users' perception of applications and data handling practices concerning COVID-19. In 2019, \citet{gerdon2021individual}, for example, conducted a CI-based longitudinal study in Germany, before the pandemic, examining people's acceptance of using individual health data during a pandemic, for public health or for private purposes. In 2020, in the wake of the pandemic, they were able to perform another such (opportunistic) study. Through the lens of CI their findings revealed that the COVID-19 pandemic altered German individuals' perspective on sharing health data with a public agency, from least acceptable before the pandemic to acceptable in the wake of the COVID-19 pandemic. Open questions remain on whether the perception will swing back after the pandemic subsides. The COVID-19 pandemic has prompted the development of ``corona apps'' for contact tracing, symptom checks, quarantine enforcement, and health certificates to help stop the spread of the virus. Using a CI-based study,~\citet{utz2021apps} examined how these applications handle health information and people's willingness to adopt them in Germany, the US, and China. They found that participants from Germany and the US perceived sharing ``corona app'' data with law enforcement agencies as inappropriate. Nevertheless, a restrictive transmission principle (e.g., limited purpose or use) increases the overall appropriateness of information flows. Additionally, compared to Germans and Americans, Chinese respondents considered sharing unique IDs with government servers and digital health certificates overall as more acceptable, highlighting the cultural differences in social norms and privacy expectations.
Our study complements the increasing body of work to examine the perceived social norms and privacy implications around pandemic mitigation technologies by focusing on VCs.

There are increasing privacy concerns about pandemic mitigation technologies re-sharing people's personal information, such as controversies related to contact tracing data being shared with law enforcement~\cite{GermanyPolice2020, AustralianPolice2021}. Building on the insights from prior studies structured by CI~\cite{Apthorper_2018,apthorpe2019evaluating,Shvartzshnaider2016,Martin_Nissenbaum_2015}, our work focuses on assessments of appropriateness that explicitly distinguish between initial information flows (i.e., when the data subject is the sender) and the subsequent re-distribution practices (when sender is a different party from subject.) Our study draws on CI to uncover the factors that are likely to affect people’s attitudes and acceptance of re-sharing of information associated with VCs. Accordingly, our study draws on CI to compare reactions both to the initial information flows as well as to the subsequent re-sharing of VC information. The outcome we seek is a comprehensive understanding of people's attitudes towards the complicated information sharing practice associated with VCs.

\section{Study Methodology}
%study goals
Our study explores the privacy and societal implications of information flows resulting from the use of vaccination certificates (VCs) in enforcing vaccination mandates (VMs).
We survey a demographically-stratified US sample on Prolific~\cite{Prolific} to investigate how various VC information sharing practices affect people's perceptions of norms.
\subsection{CI-Based Vignette Survey}\label{sec:method:ci}

%As described in Section~\ref{sec:CI_def}, the CI theory defines privacy as an appropriate exchange of information in accordance with established contextual norms. A potential privacy violation occurs when an information flow deviates from a norm. To assess potential privacy violations---i.e., when an information flow breaches the norm---in a given context, CI provides a five-parameter framework to capture information flows and norms: sender, recipient, subject of the information, information type (attribute), and the constraint on the information flow (transmission principle). When parameters' values do not align with those of an established information flow norm, then privacy is potentially violated, which requires a closer examination.

We use a CI-based vignette survey method~\cite{Shvartzshnaider2016, apthorpe2019evaluating} to gauge the effects of contextual factors on the perceived appropriateness of information sharing practices associated with common VC usage scenarios. We generated vignettes using the five CI parameters (see Table~\ref{tab:vignettes} and Figure~\ref{fig:second-hand}), based on a review of existing VC proposals~\cite{IsraelPass,Excelsior,EUGreenCert,CommonPass} and related news articles~\cite{AustralianPolice2021,Kofler2021TenReasons,GermanyPolice2020,FranceVaccinePassport,Excelsior,EUGreenCert,Martichoux2021, CAGOV2021, BritainBarVCPassport, Kiesnoski2021, Howard2021}. 
Our study included vignettes describing two types of VC information sharing practices: 
(1)``first-hand'' VC information sharing, where the sender shares their own VC information, and (2) VC information re-sharing, where the sender shares someone else's VC information. These hypothetical vignettes reflect a wide range of real-world scenarios regarding the use of VCs.

\subsubsection{First-hand information sharing  vignettes}\label{sec:first-hand} 
Using the following template with the CI parameters in Table~\ref{tab:vignettes}, we generated 21 vignettes describing ``first-hand'' information sharing when people present their VCs, as~\textit{de facto} passports, to gain access or use services potentially on a regular basis: 
\begin{framed}
\verb+[Recipient]+ ask \verb+[Sender]+ to show their
\verb+(Subject)+ vaccination certificates \verb+(Attribute)+ to \verb+[Transmission Principle]+. Would such a practice be acceptable?
\end{framed} 
To avoid potential respondent fatigue~\cite{RespondentFagitue2008Lavrakas,WebSurveyLengthQuality2009Galesic} and limit survey completion time, we presented each participant with three randomly selected vignettes out of the 21.
In addition, we curated another nine ``first-hand'' \emph{VC mandate vignettes} pertaining to in-person education, employment, international travel, and apartment rental. These nine vignettes, shown at the bottom half of Table~\ref{tab:vignettes}, are based on relevant and/or debated contexts where people comply with a VM by sharing their VCs~\cite{CAGOV2021, Kelleher2021, Faircloth2021, HalfEmployerVM, NYCEmploymentVM,FloridaLandlordVM,ApartmentVMCanada,MAAssistedLivingVM}. We showed these nine VC mandate vignettes to all participants in randomized order with an attention check.

\subsubsection{VC information re-sharing vignettes}\label{sec:second-hand} 

\begin{center}
\begin{figure*}[h]
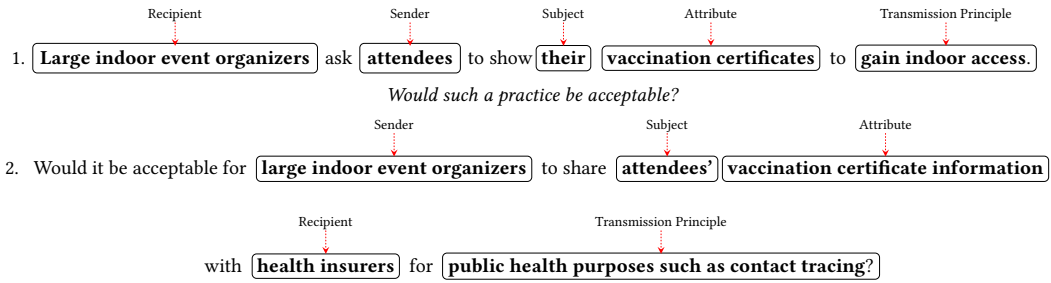

\centering
\scalebox{0.84}{
    \begin{minipage}{\textwidth}
    \centering
    %\captionsetup[figure]{name=Figure}%, label=dep:srl}
    \begin{dependency}[edge style={red,densely dotted},text only label]
        \begin{deptext}
        1. \&[0.1cm] {\bf Large indoor event organizers}  \&[0.1cm] ask \&[0.1cm] {\bf attendees}  \&[0.1cm] to show   \& {\bf their} \&[0.2cm] {\bf vaccination certificates}  \&[0.1cm] to  \&[0.1cm] {\bf gain indoor access}.\\
        \end{deptext}
        \deproot[inner sep=.5ex, edge height=5ex]{2}{Recipient}
        \deproot[inner sep=.5ex, edge height=5ex]{4}{Sender}
        \deproot[inner sep=.5ex, edge height=5ex]{6}{Subject}
        \deproot[inner sep=.5ex, edge height=5ex]{7}{Attribute}
         \deproot[inner sep=.5ex, edge height=5ex]{9}{Transmission Principle}
       % \deproot[edge unit distance=1ex]{2}{V}
    %    \depedge[edge unit distance=3ex]{3}{4}{ARG1}
     %   \depedge[edge unit distance=1.5ex]{5}{7}{ARGM-TMP}
    \wordgroup[inner sep=.3ex]{1}{2}{2}{recipient}
    \wordgroup[inner sep=.3ex]{1}{4}{4}{sender}
    \wordgroup[inner sep=.05ex]{1}{6}{6}{subject}
      \wordgroup {1}{7}{7}{attribute}
      \wordgroup[inner sep=.1ex]{1}{9}{9}{TP}
        %\wordgroup{1}{3}{4}{arg1}
        \end{dependency}
        \newline
        \textit{Would such a practice be acceptable?}
  \medskip\\
       \begin{dependency}[edge style={red,densely dotted}, text only label]
        \begin{deptext}
              2. \&[0.1cm] Would it be acceptable for \&[0.1cm] {\bf large indoor event organizers}  \&[0.1cm] to share  \&[0.1cm] {\bf attendees'} \&[0.1cm] {\bf vaccination certificate information}  \\
        \end{deptext}
              \deproot[inner sep=.5ex, edge height=5ex]{3}{Sender}
        \deproot[inner sep=.5ex, edge height=5ex]{5}{Subject}
        \deproot[inner sep=.5ex, edge height=5ex]{6}{Attribute}
        
            \wordgroup[inner sep=.05ex]{1}{3}{3}{sender}
    \wordgroup[inner sep=.05ex]{1}{5}{5}{subject}
    \wordgroup[inner sep=.05ex]{1}{6}{6}{attribute}
        \end{dependency}   
        \newline
        \newline
          \begin{dependency}[edge style={red,densely dotted}, text only label]
           \begin{deptext}
        \& with  \&[0.1cm] {\bf health insurers}  \&[0.1cm] for  \&[0.1cm] {\bf public health purposes such as contact tracing}?\\
        \end{deptext}

        \deproot[inner sep=.5ex, edge height=5ex]{3}{Recipient}
         \deproot[inner sep=.5ex, edge height=5ex]{5}{Transmission Principle}

      \wordgroup {1}{3}{3}{recipient}
      \wordgroup[inner sep=.1ex]{1}{5}{5}{TP}
        \end{dependency}

    \end{minipage}
}%
 \caption{Example of first-hand sharing (top) and re-sharing (bottom) of VC information vignette questions with marked CI parameters. Note that, as per CI theory, in the re-sharing template, the sender value does not match the subject, indicating that the sender is not sharing their own information. }\label{fig:fis-sis}
\Description{This image shows two vignettes marked with respective CI parameters. The top one reads: large indoor event organizers (recipient) ask attendees (sender) to show their (subject) vaccination certificates (attribute) to gain indoor access (transmission principle). Would such a practice be acceptable? The bottom one reads: Would it be acceptable for large indoor event organizers (sender) to share attendees' (subject) vaccination certificate information (attribute) with health insurers (recipient) for public health purposes such as contact tracing (transmission principle)?}
\end{figure*}
\end{center}

To analyze the perceptions towards possible VC information re-sharing outside the context of the original collection, we used the following question template:
\begin{framed}
 Would it be acceptable for \verb+[Sender]+ to share \verb+[Subject]+ \verb+[Attribute]+ with \verb+[Recipient]+ for \verb+[Transmission Principle]+?
\end{framed}
For the sender values in the above question template, we used the recipient values from the first-hand VC information sharing vignettes, listed in Table~\ref{tab:vignettes}, alongside additional CI parameter values in Figure~\ref{fig:second-hand}. Figure~\ref{fig:fis-sis} shows an example of the two types of vignette questions presented to participants.  

% Please add the following required packages to your document preamble:
\begin{table*}[!htb]
\centering
\resizebox{0.85\textwidth}{!}{
\begin{tabular}{l|l|l}
\bottomrule
%\hline

 \multicolumn{3}{c}{ \cellcolor{gray!20} \large{VC Passport Vignettes} }             \\
 \toprule
\textbf{Sender}     & \textbf{Recipient}   & \multicolumn{1}{l}{\textbf{Transmission Principle}}                     \\
\hline
\multirow{6}*{customers}  & restaurants and cafes                                                       &                                             \\ 
  & stores, malls, and supermarkets         &                                             \\ 
  & gyms                                                                        &                                             \\ 
  & entertainment establishments (e.g., movie theatres, museums, theatre halls) &                                             \\ 
  & personal care businesses (e.g., nail salons, barber shops)                  &                                             \\ 
  & hotels and short-term rentals (e.g., airbnb)                                &                                             \\ 
\cline{1-2}
\multirow{5}*{visitors}  & government facilities (e.g., DMVs, courthouses)                             &    \\ 
  & assisted living facilities           &    \\ 
  & hospitals          \\ 
    & places of worship                                                           &                                 \\ 
& apartment building management          &           \\ 
 & schools (K-12 and higher education)    &          \\ 
\cline{1-2}
employees  & workplaces      &   \multirow{-13}*{gain indoor access}  \\
\cline{1-3}
%\hline
\multirow{2}*{attendees}  & large indoor event organizers        &      \\ 
& large outdoor event organizers                                               & \multirow{-1}{*}{gain access}        \\ 
\cline{1-2}
\multirow{7}*{passengers} & public transportation operators       &          \\
\cline{3-3}
  & long-distance bus or train companies (e.g., Megabus and Amtrak)  & \multirow{3}*{board} \\
& cruise companies                                                            & \multicolumn{1}{l}{}  \\ 
& airline companies                                                            & \multicolumn{1}{l}{}  \\ 
\cline{3-3}
 & taxi drivers, or ridesharing drivers (e.g., Uber drivers)                     & \multirow{2}*{use the service}  \\ 
 & ridesharing companies (e.g., Uber, Lyft)                                     &         \\ 
 \bottomrule
 \multicolumn{3}{c}{\cellcolor{gray!20} \large{VC Mandate Vignettes} }               \\
 \toprule
 %\cline{2-3}
  passenger  & airlines  & take an international flight \\\hline
 foreign travelers &          & enter the United States         \\ 
US nationals      &   \multirow{-2}*{customs and border controls}        & enter a foreign country  \\ \hline
students          &   &      \\ 
teachers          &  \multirow{-2}*{schools (K-12 and higher education)}        &  \multirow{-2}*{return to in-person learning} \\  \hline
  &   & be considered for a job\\ 
   &   & be considered for a job in hospitals \\ 
  \multirow{-3}*{job applicants}  & \multirow{-3}*{employers} & be considered for a job in assisted living facilities \\ \hline
 potential renters & building management & rent an apartment\\
\bottomrule
\end{tabular}}
\caption{CI parameters used for all vignettes involving first-hand VC information sharing}\label{tab:vignettes}
\end{table*}
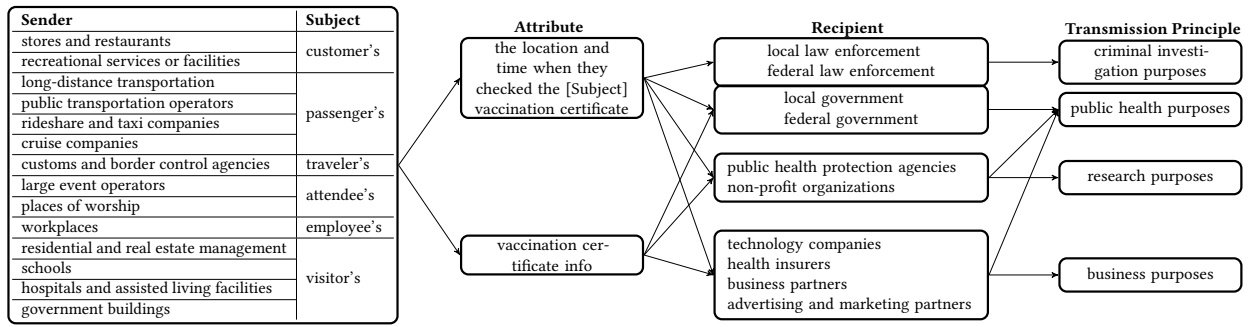
\begin{figure*}
\centering
\resizebox{0.95\textwidth}{!}{%
\centering
\tikzset{
    state/.style={
           rectangle,
           rounded corners,
           draw=black, very thick,
           minimum height=2em,
           inner sep=2pt,
           text centered,
           },
}

\begin{tikzpicture}[->,>=stealth']

 % Position of QUERY 
 % Use previously defined 'state' as layout (see above)
 % use tabular for content to get columns/rows
 % parbox to limit width of the listing

\node[state] (SENDER) 
 {\begin{tabular}{l|l}
\textbf{Sender}                                       & \textbf{Subject}                      \\ \hline
stores and restaurants                         & \multirow{2}{*}{customer's}  \\ \cline{1-1}
recreational services or facilities          &                              \\ \hline
long-distance transportation                  & \multirow{4}{*}{passenger's} \\ \cline{1-1}
public transportation operators              &                              \\ \cline{1-1}
rideshare and taxi companies                 &                              \\ \cline{1-1}
cruise companies                             &                              \\ \hline
customs and border control agencies          & traveler's                   \\ \hline
large event operators                        & \multirow{2}{*}{attendee's}  \\ \cline{1-1}
places of worship                            &                              \\ \hline
workplaces                                   & employee's                   \\ \hline
residential and real estate management       & \multirow{4}{*}{visitor's}   \\ \cline{1-1}
schools         &                              \\ \cline{1-1}
hospitals and assisted living facilities     &                              \\ \cline{1-1}
government buildings &                              \\
\end{tabular}%
};

 % State: When&Where
 \node[state,    	% layout (defined above)
  text width=10.5em, 	% max text width
  yshift=5.25em, 		% move 2cm in y
  right of=SENDER, 	% Position is to the right of QUERY
  node distance=21em, 	% distance to QUERY
  anchor=center] (WW) 	% position relative to the center of the 'box'
 {the location and time when they checked the [Subject] vaccination certificate};
 %Attribute Tag
 \node[
 text width=7.5em,
 yshift=0.675em,
 right of=SENDER,
 node distance=1.5em, 	% distance to QUERY
 anchor=center
 ] at (WW.north) 
{\textbf{Attribute}};
  % STATE VC-info
 \node[state,
  below of=WW,
  yshift=-7.5em,
  anchor=center,
  text width=10.5em] (VC) 
 {
vaccination certificate info %rmation
 %\begin{tabular}{l}
 % \textbf{QueryRep}\\
 % \parbox{2.8cm}{Dekrementiere Slotzähler}
 %\end{tabular}
 };

 % STATE Recipient-law
 \node[state,
  right of=WW,
  yshift=0.95em,
  node distance=18em,
  text width=16em,
  anchor=center] (RECIPIENT-LAW) 
 {%
 \begin{tabular}{l}
  local law enforcement\\
  federal law enforcement\\
 \end{tabular}
 };
 %Recipient Tag
 \node[
 text width=12em,
 yshift=0.45em,
 right of=WW,
 node distance=3.6em, 	% distance to WW
 anchor=center
 ] at (RECIPIENT-LAW.north) 
{\textbf{Recipient}};

  % STATE Recipient-gov
 \node[state,
  right of=WW,
  yshift=-1.9em,
  node distance=18em,
  text width=16em,
  anchor=center] (RECIPIENT-GOV) 
 {%
 \begin{tabular}{l}
  local government\\
  federal government\\
 \end{tabular}
 };

 % STATE Recipient-research
 \node[state,
  right of=WW,
  node distance=18em,
  yshift=-6em,
  text width=16em,
  anchor=center] (RECIPIENT-RES) 
 {%
 \begin{tabular}{p{15em}}
  public health protection agencies\\
  non-profit organizations\\
 \end{tabular}
 };
 
 % STATE Recipient-RES
 \node[state,
  right of=VC,
  node distance=18em,
  yshift=-1.2em,
  text width=16em,
  anchor=center] (RECIPIENT-BUS) 
 {%
 \begin{tabular}{p{15em}}
  technology companies\\
  health insurers\\
  business partners\\
  advertising and marketing partners\\
 \end{tabular}
 };

  % STATE TP-CI
 \node[state,
  right of=RECIPIENT-LAW,
  node distance=18em,
  yshift=0em,
  text width=10.5em,
  anchor=center] (TP-CI) 
 {%
 criminal investigation purposes
 };
 %TP Tag
 \node[
 text width=12em,
 yshift=0.6em,
 right of=TP-CI,
 node distance=0.9em, 	% distance to QUERY
 anchor=center
 ] at (TP-CI.north) 
    {\textbf{Transmission Principle}}; 
    
 %STATE TP-RP
  \node[state,
  right of=RECIPIENT-RES,
  node distance=18em,
  yshift=0em,
  text width=10.5em,
  anchor=center] (TP-RP) 
 {%
 research purposes
 };
  %STATE TP-BP
 \node[state,
  right of=RECIPIENT-BUS,
  node distance=18em,
  yshift=0em,
  text width=10.5em,
  anchor=center] (TP-BP) 
 {%
 business purposes
 };
 
 %STATE TP-PH
 \node[state,
  right of=RECIPIENT-GOV,
  node distance=18em,
  yshift=0em,
  text width=10.5em,
  anchor=center] (TP-PH) 
 {%
 public health purposes
 };
 % draw the paths and and print some Text below/above the graph
 \path (SENDER.east) 	edge[anchor=center]  node[anchor=north,above]{} (WW.west)
 (SENDER.east)     	    edge[anchor=center] node[anchor=north,above]{} (VC.west)
 (WW.east)       	    edge[anchor=center] node[anchor=north,above]{} (RECIPIENT-LAW.west)
 (WW.east)       	    edge[anchor=center] node[anchor=north,above]{} (RECIPIENT-GOV.west)
 (WW.east)       	    edge                               (RECIPIENT-RES.west)
 (WW.east)       	    edge                               (RECIPIENT-BUS.west)

 (VC.east)       	    edge                               (RECIPIENT-BUS.west)
 (VC.east)       	    edge                               (RECIPIENT-RES.west)
 (VC.east)       	    edge[anchor=center] node[anchor=north,above]{} (RECIPIENT-GOV.west)
 (RECIPIENT-LAW.east)    edge                               (TP-CI.west)
 (RECIPIENT-GOV.east)    edge                               (TP-PH.west) 
 (RECIPIENT-RES.east)    edge                               (TP-RP.west)
 (RECIPIENT-RES.east)    edge                               (TP-PH.west)
 (RECIPIENT-BUS.east)    edge                               (TP-BP.west)
 (RECIPIENT-BUS.east)    edge                               (TP-PH.west);
\end{tikzpicture}
}%
\caption{CI parameters used for vignettes involving re-sharing VC information}
\label{fig:second-hand}
\end{figure*}

\subsubsection{Free-text Questions} 
We asked participants additional questions about their attitudes related to COVID-19 and their vaccination status, given the divided public opinion on COVID-19 vaccines and VCs in the US~\cite{Pew2021}. 
To contextualize participants' responses to the vignettes, we included optional free-text questions to allow participants to explain their choices.

\subsection{Survey Deployment}
We administered our survey using Qualtrics~\cite{Qualtrics} and ran two pilot surveys with 75 participants each in June 2021 on the Prolific platform~\cite{Prolific}. We chose Prolific because prior findings show that their participants provide high-quality data and are relatively diverse~\cite{BeyondTurk2017Peer}. We used the results from the pilots only to improve the survey questions.

For our study, we used Prolific's ``representative sample'' option to recruit a demographically-stratified sample of 1,000 participants based on the age, gender, and ethnicity of the 2015 US Census data~\cite{ProlificRepresentative}. 
The data collection took approximately four days to complete in July 2021, and the median time spent on the survey was 13 minutes. 
Out of the 1,006 respondents recruited, we rejected six low-quality submissions and compensated the remaining participants \$2.00 for completing the survey. To further ensure data quality, we excluded results from the 110 respondents who failed one of the attention questions. In total, we analyzed valid responses from 890 participants. Their reported demographics can be found in Table~\ref{table:demographics} in the Appendix. The survey study protocol was approved by the Institutional Review Board at Carnegie Mellon.

\subsubsection{Timing of the Survey}
We conducted our survey in July 2021. At the time of the study, vaccines were widely available to all adults aged over 16, and 48.3\% of the US population was fully vaccinated (55.9\% had received at least one dose)~\cite{COVIDVaccinationRate}. 
By early July 2021, the relaxed COVID-19 measures and the Delta variant had led to a resurgence of positive cases and hospitalizations. 
At the time of the survey, 
states across the US had adopted or were about to adopt widely diverging policies regarding VCs. States such as California, New York, Louisiana, and Hawaii started to use digital vaccination records~\cite{Excelsior, HawaiiVC}, whereas states like Florida and Georgia had passed a state-wide ban on digital vaccination records~\cite{floridaBan, georgiaBan}. The debate over the use of VCs or similar vaccination verification systems remains a timely and controversial topic in public discourse~\cite{ZhangBB2021Brookings}. This study should be viewed within this particular context.%, providing a detailed snapshot of people's attitudes towards adoption of VCs about 16 months into the pandemic in the US.
\subsection{Data Analysis}

In our study, we measured people's acceptance levels towards CI-based VC usage scenarios using the 5-point Likert scale and performed a qualitative analysis of the free texts about respondents' attitudes related to COVID and VCs.

\subsubsection{Acceptance Levels for VC Information Sharing Practices}
We first compiled and graphed participants' acceptance levels towards various VC usage scenarios, which provided an overall picture of the survey responses.
Then, we ran Wilcoxon and Mann-Whitney U tests, which do not assume normal distributions, to compare ordinal distributions means of first-hand sharing and re-sharing vignettes.

\subsubsection{Regression Analysis on Acceptance Levels}\label{sec:method:clmm} We constructed a regression model of the five essential CI parameters (i.e., sender, attribute, subject, recipient, transmission principle) to measure their effects on the perceived acceptance of the respective information flow these parameters define.
For the re-sharing vignettes, we set up a cumulative link mixed model~\cite{RPackageOrdinal} (CLMM), treating perceived acceptance levels as ordinal dependent variables. The CI parameters are independent variables, and every participant is treated as a random effect. The model was fitted with the adaptive Gauss-Hermite quadrature approximation with five quadrature points. The resulting model is well defined with a condition number of the Hessian less than $10^4$~\cite{RPackageOrdinal}. The re-sharing vignettes are more suitable for a regression analysis than first-hard vignettes because the CI parameter values in the re-sharing vignettes are relatively independent of each other.
\subsubsection{Analysis of the Free-text Responses}
For the qualitative analysis of free-text responses, we conducted a streamlined thematic analysis~\cite{braun2006using} of 6,230 responses. The first author open coded all free-text responses and discussed the coded data with two other authors. We discuss the resulting themes in Section~\ref{sec:attitudes}.

 \subsection{Limitations}
Our study has several limitations. 
First, similar to previous efforts in CI-based surveys~\cite{Shvartzshnaider2016, apthorpe2019evaluating, zhang2021did}, our study is limited to the information flow space defined by the CI parameter values. As we discussed in Section~\ref{sec:method:ci}, we purposefully elicited the CI parameter values from relevant news on COVID and VC deployments. These values are not comprehensive and might change as the real-world situation evolves. Future work can examine these changes. Second, our results may not be generalizable to the US population, as crowd workers recruited from Prolific may differ from the general public. 
We tried to mitigate this issue by recruiting a large demographically-stratified sample based on the US census data. Our sample has a vaccination rate of 75\% compared with the national rate of 56\% at the time of the survey~\cite{COVIDVaccinationRate}, which might induce bias in our results. Also, we only surveyed US participants, which means the results may not apply to other nations, as information norms may vary across cultures. Finally, as with all survey work, we rely on participants' self-reported data, which may be prone to biases such as social desirability bias.

%---- END OF PAGE---
\section{Results}\label{sec:results}
This section details our analysis of vignettes as discussed in Section~\ref{sec:first-hand} where individuals are asked to share their VC information with a range of entities for various purposes and under different constraints. 
%In this section, we examine the contextual aspects of VC information sharing practices through an increasingly prevalent phenomenon: A fully-vaccinated individual is asked, in different social contexts, to share their VC information with a range of entities for various purposes and under different constraints. 
%We also take into account 
As discussed in Section~\ref{sec:second-hand}, we also examine vignettes that describe the possible re-sharing of one's VC information by the receiving entity beyond the context of the original data collection. By varying the different contextual parameters across vignettes (see Section~\ref{sec:method:ci}), we can better understand the privacy expectations and converging norms regarding VC information sharing around the following research questions:

\begin{itemize}

    \item {\bf In what contexts are VC deployments and mandates perceived appropriate?} In Sections~\ref{sec:daily} and \ref{sec:occational}, we report and compare the levels of acceptance towards VC deployment and mandate under different contexts. 
    \item {\bf How does the practice of re-sharing VC information affect the perceived appropriateness?} In Section~\ref{sec:sis}, we compare the levels of acceptance of first-hand VC information sharing (when the sender is also the subject of the information) to the re-sharing of VC information (when the sender shares someone else's information). %and investigate the effect of the four CI parameters on the perceived appropriateness.
    
\end{itemize}
%==============

\begin{figure*}[htbp]
  \centering
  \includegraphics[width=\textwidth]{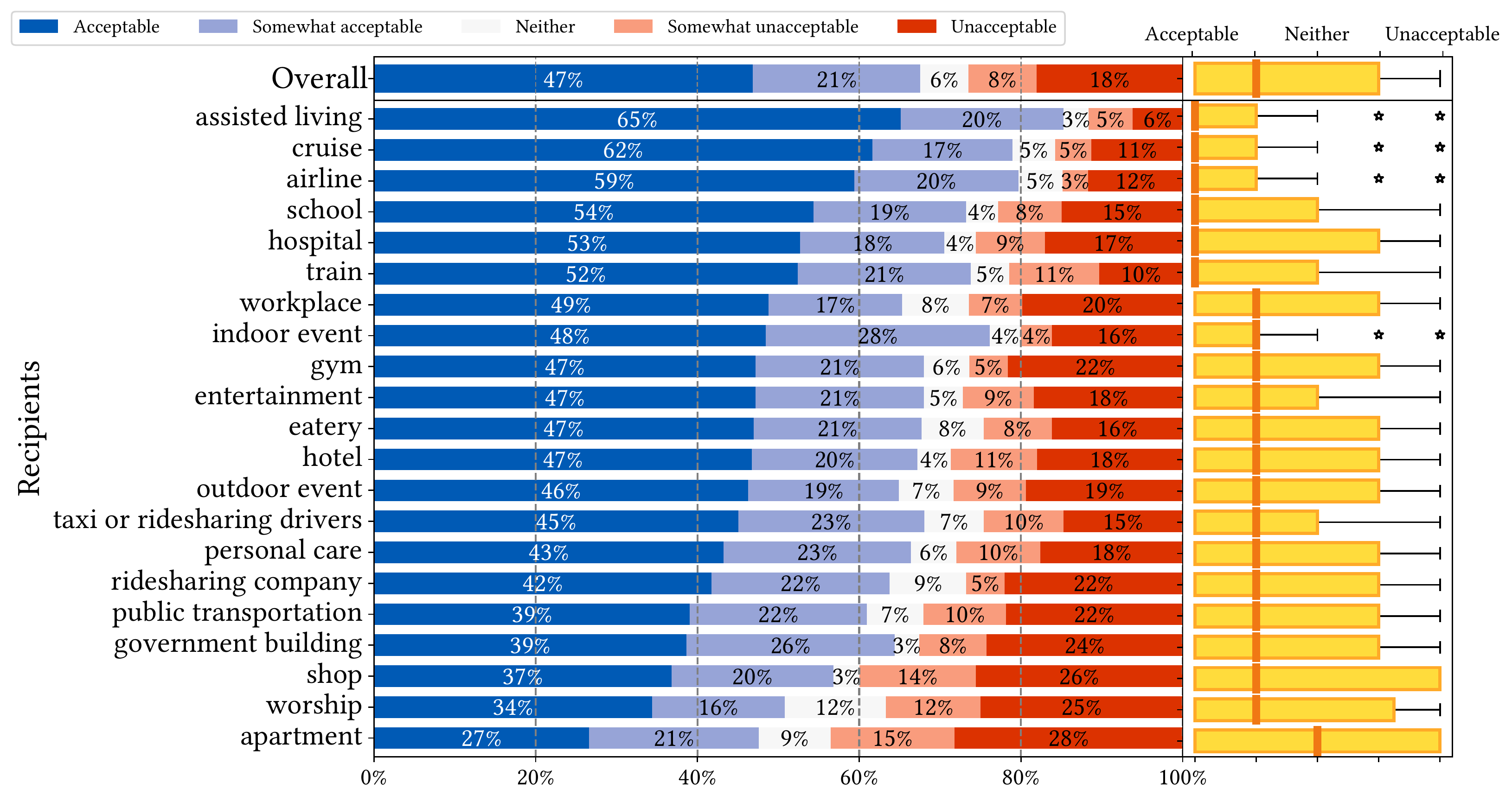}\hfill
  \caption{Reported acceptance levels for VC passport vignettes organized by recipients. The box plots on the right indicate the variances of the acceptability scores. Recall that the survey only showed each participant three randomly selected vignettes. The denominator of the percentages is the number of responses for each vignette. The top row shows an overall acceptance level across all vignettes.}
  \label{fig:place_norms}
\end{figure*} 

\subsection{VCs as \textit{de facto} passports}\label{sec:daily}
The 21 first-hand VC information sharing vignettes reflected the scenarios in which people show their VCs, as~\textit{de facto} passports, to gain access to a service, venue, or facility. 
Figure~\ref{fig:place_norms} summarizes the acceptance levels of providing VC information to 21 different CI recipients in this particular context. A majority of respondents viewed ``VC as passport" scenarios as acceptable or somewhat acceptable. For scenarios involving gaining access to assisted living facilities, cruises, and airlines, respondents expressed on average high levels of acceptance, where over 75\% of participants considered those at least somewhat acceptable. The least acceptable scenarios involve asking visitors to show their VCs to enter apartment buildings or visit worship places. Fewer than 50\% of responses indicate requiring VCs in apartment buildings as acceptable (27\%) or somewhat acceptable (21\%). 
Worship places elicit a similar reaction with only 34\% and 16\% of responses suggesting requiring VCs is ``acceptable'' and ``somewhat acceptable''. Furthermore, asking to show VCs in public transportation, government buildings, shops, worship places, and apartments elicited more diverse reactions.

\paragraph{Variances in perceptions}
We analyze variances in perceptions across 21 vignettes, which is an indicator of norm formation. A low variance is a sign of a relative agreement within the scenario. Figure~\ref{fig:place_norms} shows the variances of appropriateness scores among the scenarios. We observe low variances in perceptions for scenarios in assisted living facilities, cruises, airlines, and indoor events. This is in contrast to the high overall variances in perceptions associated with hospitals, workplaces, shops, worship places, and apartments. 
\subsubsection{\textbf{Essential services and basic facilities}}\label{sec:fis_necessity}

Our results showed that asking for VCs in a non-essential facility is considered significantly more appropriate than asking for VCs in an essential facility.
For example, 68\% of responses indicate it is ``acceptable'' (47\%) or ``somewhat acceptable'' (21\%) to show VCs in eateries compared with the lower 57\% (37\% ``acceptable'' and 20\% ``somewhat acceptable'') for showing VCs in stores. 
Several accompanying free-text comments potentially explained the discrepancy. P213 commented on restaurants requiring VCs: \textit{``The spaces are just too small, and the ambient air is not efficiently exchanged. This is the number one place for requiring people to be vaccinated. People are voluntarily choosing to go, so should have to show a pass.''} Yet, P156 noted: \textit{``Freedom to access a source of food such as a supermarket should be effortless. Having to show vaccination certificates to enter would cause mayhem.''} 

Noticeably, asking visitors to show their VCs in hospitals is significantly less acceptable than that in assisted living facilities (Mann–Whitney U $= 7924.5$, $n_1 = 117$, $n_2 = 116$, $p < 0.01$, $Cohen'd = 0.293$, two-tailed), although both are places with COVID-19 vulnerable populations. 
%Some participants argued that people should not be deprived the right to visit healthcare providers such as hospitals due to their vaccination status. For instance, 
P72 provided a possible line of reasoning:~\textit{``Even though some people may not feel comfortable getting the vaccine, they should still be granted access to hospital resources indoors. If an individual lacks certification, they should be wearing a mask.''}

The results also reveal a similar contrast between public transportation and other forms of transportation such as airlines, trains, and taxis or ride-sharing services. Only about 60\% of respondents found it ``acceptable''(39\%)  or ``somewhat acceptable''(22\%) to show their VCs to use public transportation. P209's open-ended comment provides some context to the reported contrast: \textit{ ``Safety is important here too, but unlike flying, public transportation is more of a necessity and shouldn't be hindered by this.''}

In summary, the results highlight the relationship between the nature of the context---whether it is deemed essential or non-essential---and the perceived appropriateness. This suggests a need for nuanced policy making with regard to using VCs as passports.

%\subsubsection{GROUP B: \textbf{Acceptance Levels for Most FIS Occasional Vignettes Are High}}~\label{sec:occational}
\subsection{Examining VC mandate vignettes}~\label{sec:occational}
\begin{figure*}[htbp]
  \centering
     \includegraphics[width=\textwidth]{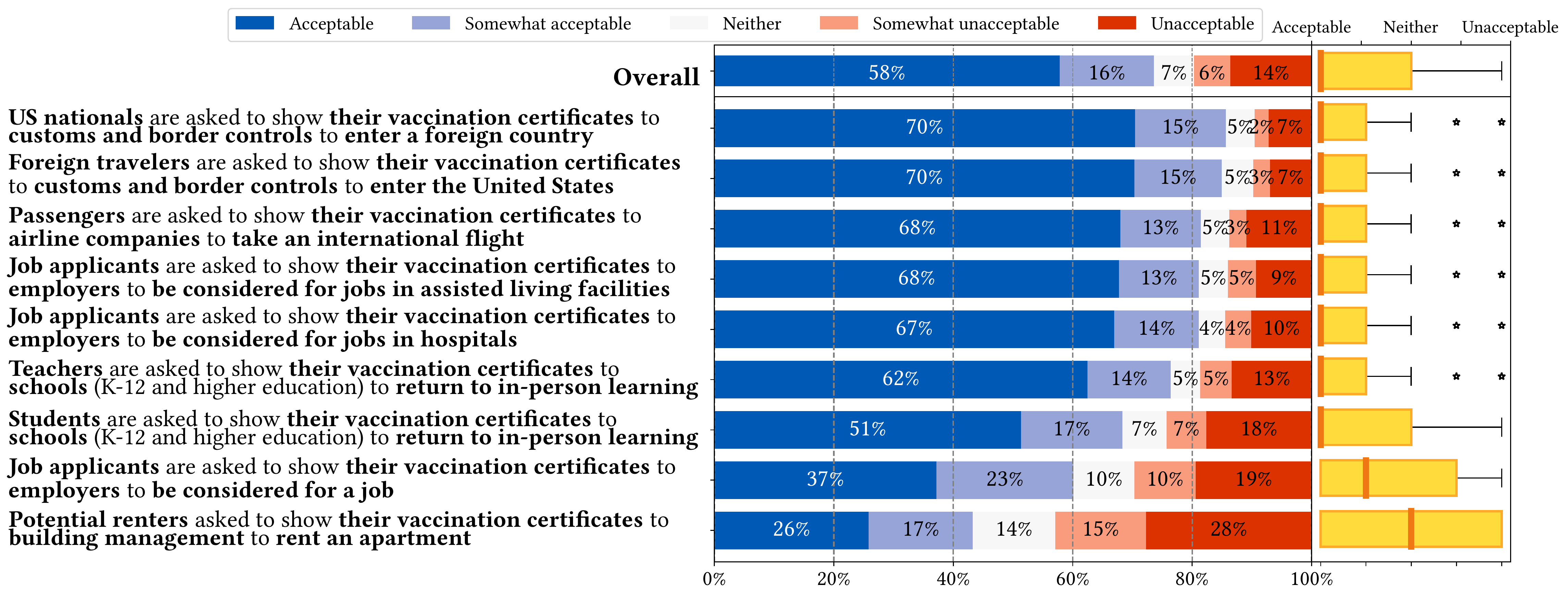}\hfill
  \caption{Participants' acceptance levels for nine vignettes. The top row displays the averaged response across nine vignettes. The right graph shows a box plot of the ordinal data with the mean marked in orange.}
  %The bottom row shows the aggregated preferences across all places.
  \label{fig:vignette_norms}
\end{figure*}

%Furthermore, Figure~\ref{fig:vignette_norms} shows on average high acceptability levels for vaccination certificates in all nine scenarios with 47\% finding them to be acceptable and 21\% somewhat acceptable. 
The nine VC mandate vignettes reflected publicly debated scenarios in the context for which governments around the world are seeking mandates to require VCs, such as for international travel, returning to in-person learning, applying for a job, and renting an apartment, as mentioned in Section~\ref{sec:first-hand}. Figure~\ref{fig:vignette_norms} summarizes the acceptance levels for each of the nine vignettes from all participants. Overall, 74\% of participants found the selected vignettes to be ``acceptable'' (58\%) or ``somewhat acceptable'' (16\%).

\paragraph{\textbf{A VC mandate for international travel is perceived appropriate to take a flight or use at the border.}}
Our results show that requesting VCs for international travel is largely perceived as appropriate: 82\% of all participants stated that it is acceptable (68\%) or somewhat acceptable (14\%) for passengers to show VCs to take an international flight. Similarly, 85\% of respondents perceived showing VCs to customs and border control agencies as acceptable (70\%) or somewhat acceptable (15\%), both for entering the US or a foreign country. 

\paragraph{\textbf{A VC mandate for employment: Perceived appropriate to apply for a job at assisted living facilities or hospitals.}}
81\% of respondents expressed similar levels of acceptability for sharing vaccination certificate information with employers to be considered for a job in assisted living facilities and hospitals, with 14\% stating it was somewhat acceptable and 67\% viewing it as acceptable. In comparison, when it comes to applying for a general position, only 60\% participants considered showing vaccination certificates to potential employers for a job as acceptable (37\%) or somewhat acceptable (23\%).  
A Wilcoxon Signed-Rank test shows that levels of acceptance for the general case were statistically significantly lower than the levels of acceptance for the cases involving hospitals and assisted living facilities ($Z=5.42,  p < 10^{-26}$).

\paragraph{\textbf{A VC mandate for education: Perceived appropriate for teachers, less so for students}}
When asked whether it is appropriate to share VC information with schools for returning to in-person learning, the acceptance levels depended on whether the sender is the students or teachers. These two vignettes involved the same CI parameters except for the sender. 
Using a Wilcoxon Signed-Rank test ($Z=9.89, p<0.000001$), we noted the perceived levels of acceptability for students were statistically significantly lower than those for teachers. This means that even though the majority of our survey respondents considered the VC mandate in schools acceptable, they regarded asking students to share their VC information with the school as less acceptable than asking teachers to do so.

\paragraph{\textbf{A VC mandate in residential settings: Perceived as inappropriate overall}}
Respondents viewed showing VCs to building management to rent an apartment as the least acceptable. With only 17\% stating that it was somewhat acceptable, a slightly higher percentage of the respondents (26\%) saw it as acceptable. Such low acceptance is also consistent Section~\ref{sec:fis_necessity} where respondents considered showing VCs to visit an apartment as the least acceptable.

%\begin{table}[htb!]
%\centering
\begin{table*}[h]

\resizebox{0.6\textwidth}{!}{
\begin{tabular}{llllp{5em}}
%\begin{tabular}{l S[table-format=2]S[table-format=2]S[table-format=2]S[table-format=2]}
%\bottomrule
\textbf{Factors}                  & \textbf{Est.} & \textbf{Std. Err} &\textbf{Z} & \textbf{$p$-value}\\ \bottomrule
%Intercept                  & -1.79965 & 0.60789 & -2.96  & 0.003072\twoS  \\%\hline
\multicolumn{5}{c}{\cellcolor{gray!20} \textbf{Sender}: baseline=customs and border control }                 \\ %\hline
\toprule
government buildings                     & 0.0448  & 0.1042   & 0.4299 & 0.6672\\ %\hline
hospitals and assisted living facilities & 0.1404  & 0.1057   & 1.3280 & 0.1842\\
long-distance transportation             & 0.1713  & 0.1022   & 1.6758 & 0.0938\\
cruise companies                         & 0.1921  & 0.1052   & 1.8256 & 0.0679\\
workplaces                               & 0.2633  & 0.1079   & 2.4404 & 0.0147\oneS\\
large event organizers                   & 0.3089  & 0.1080   & 2.8603 & 0.0042\twoS\\%\hlinehttps://www.overleaf.com/project/61a68c6ce7a699bf06ed989b
schools                                  & 0.3545  & 0.1072   & 3.3076 & 0.0009\threeS\\
stores and restaurants                   & 0.3804  & 0.1053   & 3.6127 & 0.0003\threeS\\
recreational services or facilities      & 0.4255  & 0.1062   & 4.0054 & 6.2e-5\threeS\\%\hline
public transportation operators          & 0.4280  & 0.1081   & 3.9588 & 7.5e-5\threeS\\
places of worship                        & 0.4600  & 0.1090   & 4.2195 & 2.4e-5\threeS\\
residential and real estate management   & 0.5122  & 0.1090   & 4.7005 & 2.6e-6\threeS\\%\hline
rideshare and taxi companies             & 0.6041  & 0.1046   & 5.7761 & 7.7e-9\threeS\\
\bottomrule
\multicolumn{5}{c}{\cellcolor{gray!20} \textbf{Recipient}: baseline=public health protection agencies}                \\ %\hline
\toprule
local government             & 0.9994 & 0.0839 & 11.9142 &  2.2e-16\threeS\\%0.00000000000000022\threeS \\%\hline
federal government           & 0.9788 &  0.0838 & 11.6736 &  2.2e-16\threeS\\%0.00000000000000022\threeS \\%\hline
non profit organization      & 2.2955  & 0.0813 & 28.2482 & 2.2e-16\threeS\\% 0.00000000000000022\threeS \\%\hline
health insurer               & 2.5199  & 0.0821 & 30.7003 & 2.2e-16\threeS\\% \threeS \\%\hline
business partners            & 3.9754 &  0.0872  &  45.5826 & 2.2e-16\threeS\\%\threeS \\%\hline

technology company           & 4.2075 & 0.0886 & 47.4724 & 2.2e-16\threeS\\%\threeS \\%\hline
advertising and marketing partners & 4.7668 &  0.0926  &  51.4647 & 2.2e-16\threeS\\ \bottomrule%\hline%\threeS \\ 
\multicolumn{5}{c}{\cellcolor{gray!20} \textbf{Attribute}: baseline=vaccination certificate information}      \\ %\hline      
\toprule
location and time     & -0.0472  & 0.0415  &  -1.1355 & 0.2561 \\ 
\bottomrule
\multicolumn{5}{c}{\cellcolor{gray!20} \textbf{Transmission Principle}: baseline=public health purposes}                           \\ %\hline
\toprule
criminal investigation     &0.7099 & 0.0819 & 8.6722 & 2.2e-16\threeS\\%0.00000000000000022\\ 
research                   &0.6427 & 0.0927 & 6.9334 & 4.1e-12\threeS\\%0.00000000000000022\\ 
business                   &0.2994 & 0.0475 & 6.3036 & 2.9e-10\threeS\\%0.00000000000000022\\ 

\bottomrule
\end{tabular}}
\caption{Cumulative Linear Mixed Model Regression. A positive coefficient (estimate) shows participants' decreased acceptance}\label{tab:clmm_tbl2}
\end{table*}

\subsection{Examining Scenarios on Re-sharing VC Information}\label{sec:sis}

We examined respondents' perceived appropriateness regarding VC information re-sharing practices: a situation in which a VC shown in a given context is being shared by the original recipient with a different entity for a new purpose or under a new condition. For example, when businesses share their customers' VC information with the health protection agency for public health purposes such as contact tracing. For a full list of vignettes and CI parameters, see Figure~\ref{fig:second-hand}. 

\begin{figure*}[htbp]
  \centering
  \includegraphics[width=\textwidth]{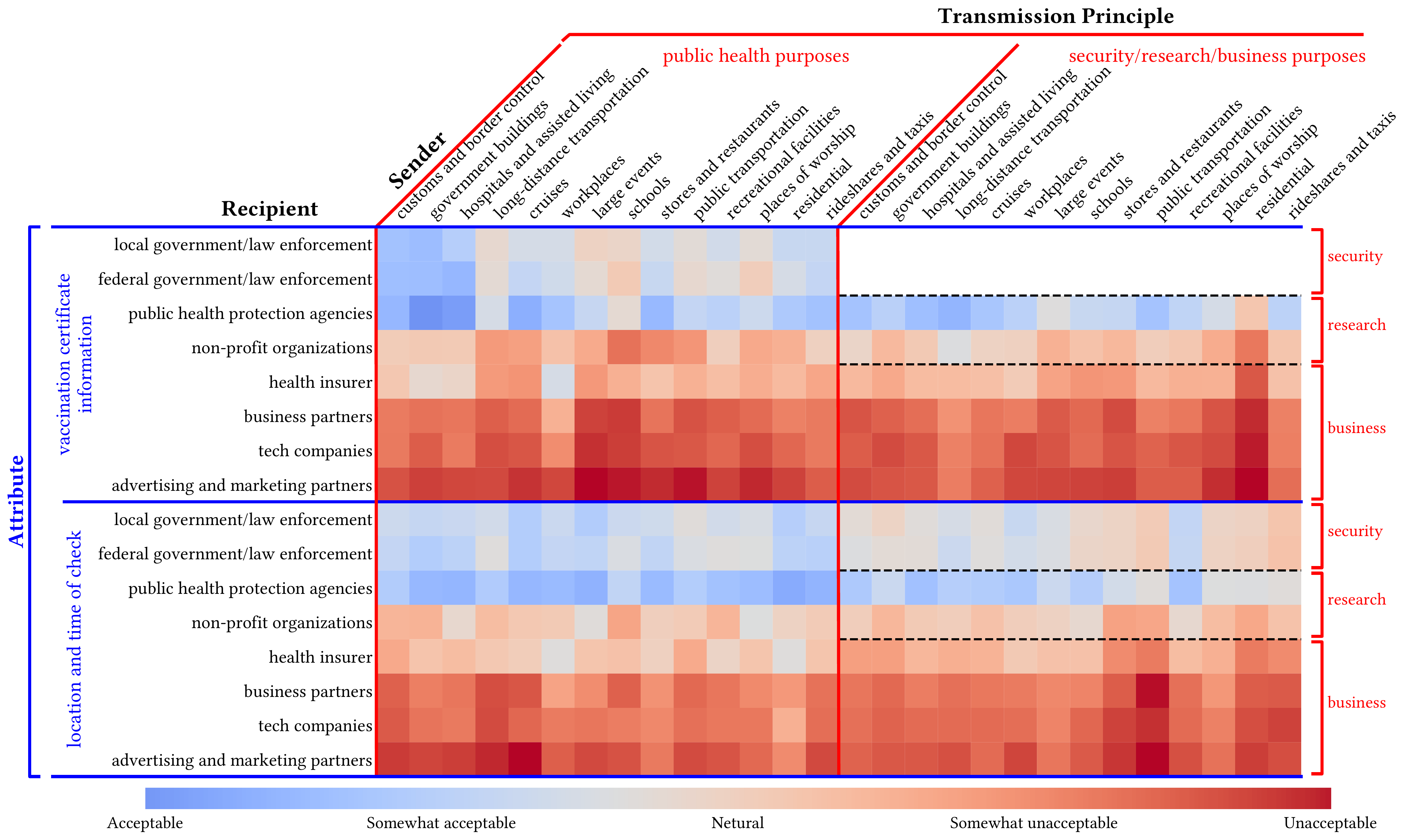}\hfill
  \caption{A heat map of the average of all participants' responses under a combination of four CI parameters (sender, recipient, attribute, and transmission principle). For instance, the color of the top left cell represents the acceptance level of the information flow---customs and border control agencies share their customers' vaccination certificate information with the local government for public health purposes.  }\label{fig:table2_2row}
  \Description{The x-axis of the heat map lists the different senders and transmission principles. The y-axis of the heat map lists the various recipients and attributes. The color red indicates unacceptable, whereas the color blue indicates acceptable. Sharing VC information with public health protection agencies for public health purposes seems to be the most acceptable. }
\end{figure*}

%\paragraph{\textbf{Re-sharing of VC-related information is considered less acceptable}}
Figure~\ref{fig:table2_2row} shows a heat map of average acceptance levels of vignettes describing VC information re-sharing. Overall, the practice of re-sharing and re-purposing of VC information is perceived as less appropriate compared with the first-hand VC information exchange in the original context. We found a statistically significant difference between the two types of information flows using a Wilcoxon matched-pair signed rank test ($Z=4.80, p<10^{-7}$).

\subsubsection{Regression analysis of vignettes' CI parameters}
A closer examination of CI parameters in the re-sharing vignettes reveals varied levels of perceived appropriateness. 
Table~\ref{tab:clmm_tbl2} shows the results of the CLMM regression analysis (see Section~\ref{sec:method:clmm}) of factors affecting participants' perceived acceptance levels of re-sharing VC information. We found that values of three CI parameters---sender, recipient, and transmission principle---have a statistically significant effect on participants' perceived appropriateness.

\paragraph{Sender}
We used ``customs and border control agencies'' as the baseline in our regression analysis of the sender parameter as such a sender is the most accepted among all senders. 
Out of all 14 senders of the vignettes, rideshare drivers/companies and residential management were perceived as the most unacceptable sender, with respective odds ratios of 1.6 (=$e^{0.6041}$) and 1.4 (=$e^{0.5122}$) compared to the baseline value. In other words, VC information sharing by rideshare companies is 1.6 times less acceptable than customs and border controls, holding constant all other variables. 
%As illustrated in the box plot in Figure~\ref{fig:ciparam}, re-sharing by residential places, places of worship, schools, and public transportation operators elicited a median response of unacceptable.

\paragraph{Recipient}

Our results show that sharing VC information with public health protection agencies (which we used as the baseline) is significantly more acceptable than sharing VC information with other receiving entities. The least unacceptable recipients included advertising and marketing partners, followed by technology companies, and business partners, as indicated by the decreasing coefficients.  %illustrated in the box plot in Figure~\ref{fig:ciparam}. 

\paragraph{Transmission Principle}
Our results indicate that sharing VC-related information for public health purposes (the baseline) is significantly more acceptable than for other purposes or conditions. % values.  

\paragraph{Attribute}
In addition to the information in the VC itself, we looked at the meta-data associated with VC information sharing, such as the location and the timestamp. This information, when shared with other entities, could be further used to surveil or track individuals. 
Our analysis, however, shows no statistically significant difference in perceptions of re-sharing the VC information or the residual meta-data (location and time) associated with the VC check.

\begin{comment}
\begin{figure*}
    \centering
        \begin{subfigure}[b]{0.495\textwidth}
            \centering
                        \includegraphics[width=\textwidth]{figures/table2_sender.pdf}
            \caption[]%
            {{\small Sender}}    
            \label{fig:ci:sender}
        \end{subfigure}
        \hfill
        \begin{subfigure}[b]{0.495\textwidth}  
            \centering 
            \includegraphics[width=\textwidth]{figures/table2_recipient.pdf}
            \caption[]%
            {{\small Recipient}}    
            \label{fig:ci:recipient}
        \end{subfigure}
        \caption[ The average and standard deviation of critical parameters ]
        {\small Box plots of perceived appropriateness organized by values of different CI parameters} 
        \label{fig:ciparam}
    \end{figure*}
\end{comment}

\paragraph*{\textbf{Summary}}
Our analysis shows that contextual factors captured by the CI framework affect the degree to which participants judged a VC practice acceptable. Some combinations of sender, subject, recipient of the VC information and the condition/constraint of the transfer (transmission principle) have a statistically significant effect on the perceived acceptance of the information flow that these parameters define. This aligns with prior work that leverages CI to evaluate privacy violations in other contexts~\cite{Shvartzshnaider2016, apthorpe2019evaluating}. Notably, the \verb|subject| parameter of the information flow is particularly important, as it distinguishes re-sharing practices. Our participants found VC-related information re-sharing practices less acceptable than their providing VC directly to recipients.

\subsection{Different Views on VCs: Qualitative Analysis}\label{sec:attitudes}

The open-ended comments accompanying the vignettes provide insight into the motivating factors behind the stated perceptions of appropriateness. Our thematic analysis of the free texts reveals three main attitudes. %We present the identified themes per attitude. 

\subsubsection{In favor of VCs}
%We present rationales of participants who support VC usage.
Over half of participants (57\%) noted that VCs would make them feel safer or curb the spread of the virus. %, including dangerous variants.
Some (12.5\%) also mentioned VCs can show proof and prevent counterfeit CDC cards, or carrying digital VCs are easy and safe from losing them. 7.6\% of participants referred to communitarian ethics in helping protect others and their community, while others (3.8\%) saw no difference from existing practices like showing IDs to buy alcohol. 2.5\% of participants commented that VCs would serve as an incentive to get more people vaccinated. 

\subsubsection{Opposing VCs}
Some (11.7\%) participants indicated they regard VCs as an invasion of their privacy or, more generally, a restriction on their personal freedom. Others (7.3\%) believed that deciding on whether to receive the vaccine should be left to their individual discretion instead of being imposed by the government or some other organization. 6.6\% of participants considered vaccination status as private medical information and thus information that should not be shared with anyone other than their doctors. 4.3\% claimed that VCs are illegal and/or unconstitutional and/or violated their HIPAA rights. 
Concerned about potential harms, some (6.2\%) perceived VCs as a form of government overreach, compared the practice to identity control measures such as the ones under the Nazi regime (``Papers, please''). 

A few participants (4.7\%) referred to information privacy, indicating that they were not comfortable with some information included in VCs or would not want such data retained or shared. Some (4.0\%) noted that employing VCs would result in discrimination against the unvaccinated. 
%\aerin{add summary}

\subsubsection{Context-sensitive views}
37\% of participants expressed mixed reactions and considerations dependent on the contexts of VC information sharing. 
For example, 11.2\% thought that private businesses are free to require VCs at their own discretion, and 7.6\% were against requiring VCs at places to which people need access, such as public transportation and stores. 5.7\% of participants believed that other methods such as mask wearing, negative COVID tests, and occupancy limits should also be accepted if some would not want to present their VCs. 4.6\% of participants mentioned the need to accommodate people who may not be able to receive vaccines when deploying VCs. 3.9\% of participants thought that VCs are particularly controversial and could elicit strong objections and potentially violent behaviors.

\section{Discussion}
As we write this paper, VMs and VCs remain a highly contentious and politically polarizing subject. Faced with the new and highly infectious omicron variant, many governments around the world have introduced vaccination mandates or the use of vaccination certificates across a number of different contexts~\cite{IsraeliWorkplace,CanadaVaxpass,APAustrian}.
The intensity and polarization of the debate is vividly reflected in the views expressed by the participants in our study. At the one extreme, a handful of participants left profanities in the free-text responses, aimed at the authors whom they mistakenly thought were conducting research to shore up support for VC mandates and deployments. At the other extreme, a few participants left equally strong responses about people's collective responsibility to protect one another, asserting that those who refuse vaccinations are selfishly neglecting their responsibility. 

Aside from the extremes, at an aggregate level, the percentage of people who find appropriate many of the VC sharing scenarios presented to them, could be taken as potential support for a fairly broad VC mandate. A closer look, however, reveals a more nuanced picture in which contextual factors had significantly affected participants' attitudes. It mattered whether the VC information is shared with the school to facilitate in-person classroom, with a grocery store owner or with a gym operator as a condition of admittance, or with a customs agent to enter a country. The recipient with whom VC data is shared, the purpose(s) for sharing, as well as guarantees (or lack thereof) about the processing of VC information all have a significant effect on people's acceptance of VC deployments. It is worth noting, too, that our study found the \verb|subject| parameter of the information flow to be important, lending credence to our initial question about first-hand use versus re-sharing practices. 
When the values for all the parameters are clearly stated, our results indicate a negative sentiment towards requiring VCs for access to essential services and activities, places of worship, and apartment buildings. Further, perhaps not surprisingly, the practice of re-sharing VC information is perceived as largely inappropriate. These empirical results illustrate the importance of organizing a survey like this one by systematically sampling different contextual values, especially when it comes to understanding people's acceptance of information flows associated with different possible VC deployments and their implications. 

Finally, as posited by the CI theory~\cite{nissenbaum2014respect}, newly-formed information flows that challenge established norms can affect the ultimate realization of a range of societal values such as equality, equity, and civil liberties. The assessment of the appropriateness of new flows includes: 1) a cost and benefit analysis of the information flow related to all the affected parties: Who benefits? What risks are involved?  2) a review of moral and ethical values such as fairness, autonomy, and informational harm; 3) considerations around how the new information flow contributes to fulfilling the ``context-specific values, ends and purposes''~\cite{nissenbaum2014respect}. 

The qualitative analysis of the open-ended responses in Section~\ref{sec:attitudes} reveals that the ethical and societal values indeed are part of the normative assessment of the perceived appropriateness of VCs. The open-ended comments included different aspects related to the appropriateness assessment. We observed the weighing of public health interests against the expectations of freedom and privacy in various contexts. Many participants reported viewing enhanced public health as a societal benefit, while some were concerned about potential harm brought by heightened government surveillance. Some participants also expressed concerns about their bodily autonomy, the violation of personal freedom, and the intrusion of privacy on their health information, while others also warned of potential discrimination against the unvaccinated and restrictions on their rights to access essential facilities such as stores and hospitals.

\section{Conclusion}
We presented a US-based online survey study aimed at gauging people's acceptance of VCs across a diverse collection of possible deployment scenarios. This work is unique in its recognition of the fundamental privacy questions entailed by the deployment of VCs and differs from other surveys in its use of Contextual Integrity as an organizational framework to systematically explore possible deployment scenarios and contextual parameters.
Our study illustrates how Context Integrity (CI) provides an effective framework for approaching controversial societal practices, such as VC deployment. It suggests that the multifactorial insights that CI yields can inform richer and more nuanced responses to challenges confronting society in today's fight against COVID-19, and potentially other similar challenges going forward. Our study shows that contextual parameters can significantly affect people's judements about what is and isn't appropriate in the deployment of VCs. 
In the context of vaccination mandates and certificates, beyond the blunt approach one often hears---that privacy must be traded off against public health---our findings open the door to more informed and nuanced alternatives that allow the pursuit of public health even as we reinforce appropriate information-flow practices that conform with the wide attitudes of ordinary people.

%\clearpage

%% The acknowledgments section is defined using the "acks" environment
%% (and NOT an unnumbered section). This ensures the proper
%% identification of the section in the article metadata, and the
%% consistent spelling of the heading.
\begin{acks}
This research has been supported in part by the National Science Foundation (NSF) Secure and Trustworthy Computing program (grants CNS-1801316 and CNS-1801307). The US Government is authorized to reproduce and distribute reprints for Governmental purposes, notwithstanding any copyright notice thereon. The views and conclusions contained herein are those of the authors and should not be interpreted as representing the official policies or endorsements, either expressed or implied of NSF or the US Government.
\end{acks}

%%
%% The next two lines define the bibliography style to be used, and
%% the bibliography file.
\bibliographystyle{ACM-Reference-Format}
\bibliography{bibliography}

%%
%% If your work has an appendix, this is the place to put it.
\appendix
\section{Appendix}
\subsection{Survey}\label{apx:survey}
\begin{comment}
\subsubsection{Consent Form}
\begin{itemize}
    \item I am age 18 or order.
    \item I have read and understand the information above.
    \item I want to participate in this research and continue with the survey.
\end{itemize}
\end{comment}
\subsubsection{Introduction}
With ongoing COVID-19 vaccination efforts, governments and other organizations around the world have proposed the use of “vaccination certificates” as a way to verify that a person has been vaccinated against the coronavirus, received a negative test or has recovered from the virus. Some vaccination certificates are already in use today. Researchers at Carnegie Mellon University are conducting a study to understand people’s opinions and perceptions of these vaccination certificate proposals. Please answer the survey honestly. There are no right or wrong answers to any of the questions. 
\begin{comment}
Do you agree or disagree with the following statements?
\begin{itemize}
    \item Vaccination certificates will be commonplace in the next 3 to 6 months.
    \item Vaccination certificates are useful for a safe return to normal.
    \item Would you prefer these vaccination certificates to be $\_\_\_\_$? %Please select all that apply.
\end{itemize}
\end{comment}
\subsubsection{Fist-hand Information Sharing: Vaccination Passport Vignettes}
\begin{itemize}
    \item Pre-COVID, how often did you visit \verb+[place]+?
    \item Assume that you have a vaccination certificate similar to the one below. 
    \begin{figure}[h]
        \centering
        \includegraphics[width=\columnwidth]{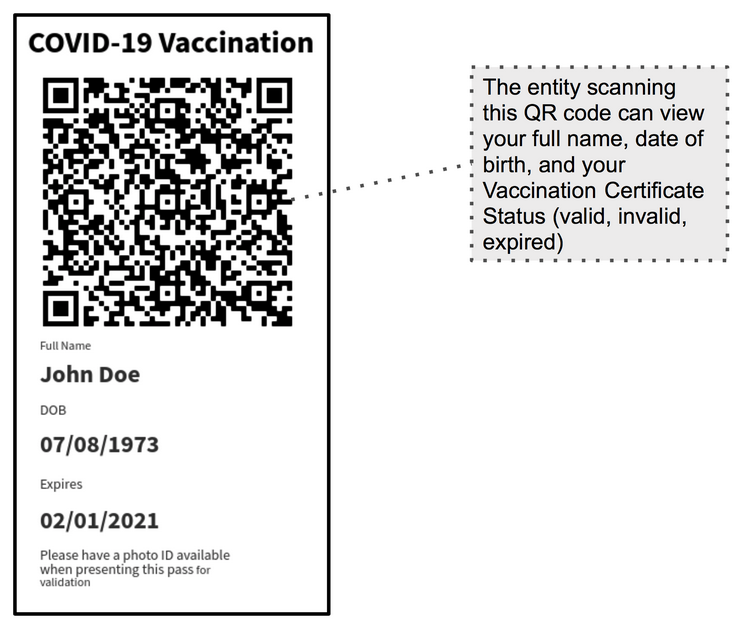}
        \caption{An example vaccination certificate shown to survey participants.}
        \Description{The image displays a COVID-19 vaccination certificate with a QR code, the holder's name listed as John Doe, his date of birth listed as July 8th, 1973, and expiration date Feb. 1st, 2021. At the bottom of the certificate, it says: Please have a photo ID available when presenting this pass for validation. There is a text box to the right of the pass that says: The entity scanning this QR code can view your full name, date of birth, and your Vaccination Certificate Status (valid, invalid, expired).}
        \label{fig:my_label}
    \end{figure}
    \item Template: \verb+[Recipient]+ ask \verb+[Sender]+ to show \texttt{[Subject +Attribute]} to \verb+[Transmission Principle]+? Would such a practice be acceptable?
    \item Example: [Gyms] ask [members] to show [their vaccination certificates] to [gain indoor access]. Would such a practice be acceptable? Please explain.
    \item If such a certificate were to be required to [gain indoor access], how much more likely would you be to go to [gyms] over the next 6 months? Please explain.
\end{itemize}
\subsubsection{Fist-hand Information Sharing: Vaccination Mandate Vignettes}
\begin{itemize}
    \item Passengers are asked to show their vaccination certificates to airline companies to take an international flight. Is this acceptable? 
    \item Foreign travelers are asked to show their vaccination certificates to customs and border controls to enter the United States. Is this acceptable?
    \item Us nationals are asked to show their vaccination certificates to customs and border controls to enter a foreign country. Is this acceptable?
    \item Teachers are asked to show their vaccination certificates to schools (K-12 and higher education) to return to in-person learning. Is this acceptable?
    \item Students are asked to show their vaccination certificates to schools (K-12 and higher education) to return to in-person learning. Is this acceptable?
    \item Job applicants are asked to their show vaccination certificates to employers to be considered for a job. Is this acceptable?
    \item Job applicants are asked to show their vaccination certificates to employers to apply for jobs in hospitals. Is this acceptable?
    \item Job applicants are asked to show their vaccination certificates to employers to apply for or retain jobs in assisted living facilities. Is this acceptable?
    \item Potential renters asked to show their vaccination certificates to building management to rent an apartment. Is this acceptable?
    \item Word count: Please select the answer choice with the largest number of words in the list below.
\end{itemize}

\subsubsection{VC Information Re-sharing Vignettes}
\begin{itemize}
    \item Template: Would it be acceptable for \verb+[Sender]+ to share \verb|[Subject Attribute]| with the following entities for \\\texttt{[Transmission Principle]}? \\ 
    \ Example: Would it be acceptable for [recreational services or facilities (e.g., bars, gyms, salons)] to share [information on a person’s vaccination certificate] with the following entities [for public health purposes such as contact tracking]?
\end{itemize}
\subsubsection{Vaccination Certificate Questions}
\begin{itemize}
\item Do you agree or disagree with the following statements?\\
The government (federal or state)
\begin{itemize}
\item should promote vaccination against COVID-19.
\item has no right to impose vaccination certificates.
\item should issue vaccination certificates and require them to be used in different contexts.
\end{itemize}
    \item Which entity do you consider trustworthy to develop a vaccination certificate? Please select all that apply. Please explain.
    \item Would you prefer to have a single certificate issued by the federal government and recognized by everyone or different certificates issued by different organizations from which you can choose?
\end{itemize}
\subsubsection{COVID Related Questions}
\begin{itemize}
    \item Have you been vaccinated against COVID-19?
    \item Have you contracted COVID-19? 
    \item Do you personally know anyone who got seriously ill due to COVID-19?
    \item If vaccination certificates were to be used, would you be more or less likely to get vaccinated?
\end{itemize}
\subsubsection{Demographics}
\begin{itemize}
    \item What is your age?
    \item What is your gender?
    \item What is the highest level of education you have completed?
    \item What was your total household income before taxes during the past 12 months?
    \item What is your marital status?
    \item Which of the following best describes your primary occupation?
    \item Please specify your ethnicity.
    \item In general, would you describe your political views as \_\_\_?
    \item Have you ever held a job in assisted living facilities or hospitals?
    \item Do you use a smartphone? 
    \item In which state do you currently reside [drop-down]?
    \item Which category best describes where you live?
    \item Have you used the following tools in the past year? Please select all that apply.
    \item Not including this survey, approximately how many surveys related to privacy or security have you completed in the past year?
    \item Anything else you’d like to say about the situation and/or your concerns? 
\end{itemize}

\subsection{Sample Demographics}
\begin{table}[hb]
\resizebox{\columnwidth}{!}{%
\begin{tabular}{lr|lr|lr}
\multicolumn{2}{c|}{\cellcolor{gray!20}Gender} & \multicolumn{2}{c|}{\cellcolor{gray!20}Age} & \multicolumn{2}{c}{\cellcolor{gray!20}Ethnicity} \\ \bottomrule
Female & 51.0\%  & 18--27 & 19.6\%  & Asian  &  6.6\% \\ 
Male   &  47.4\%  & 28--37 & 18.8\%  & African American    &  12.5\% \\
 Other &  1.2\% &  38--47  & 16.6\% & Caucasian  & 71.6\%   \\  
Decline to answer & 0.3\%  & 48--57 & 16.3\% & Hispanic &  4.7\% \\
&   &  58+ & 28.8\% & Other & 3.6\%\\   
&   &      &        & Decline to answer& 1.0\% \\ 
\bottomrule
\end{tabular}%
}
\caption{Demographics of our study participants $N=890$}
\label{table:demographics}
\end{table}

\end{document}